\documentclass[12pt]{article}

\usepackage{fullpage}
\usepackage{latexsym}
\usepackage{epsfig}

\begin{document}

\title{Dihedral angles and orthogonal polyhedra}
\author{Therese Biedl 
\thanks{David R. Cheriton School of Computer Science, University of Waterloo, 
Waterloo, ON N2L 3G1, Canada.
Research of TB and AL supported by NSERC.}
\and Martin Derka 
\addtocounter{footnote}{-1}\footnotemark
\and Stephen Kiazyk 
\addtocounter{footnote}{-1}\footnotemark
\and
Anna Lubiw 
\addtocounter{footnote}{-1}\footnotemark
\and Hamide Vosoughpour
\addtocounter{footnote}{-1}\footnotemark
}
\date{\today}

\maketitle

\section{Introduction}

Consider an orthogonal polyhedron, i.e., a polyhedron where (at
least after a suitable rotation) all faces are perpendicular to a coordinate
axis, and hence all edges are parallel to a coordinate axis.  Clearly,
any {\em facial angle} (i.e., the angle of a face at an incident vertex)
is a multiple of $\pi/2$.  Also, any {\em dihedral angle} (i.e., the
angle between two planes that support to faces with a common edge) is
a multiple of $\pi/2$.

In this note we explore the converse: if the facial and/or dihedral angles are all multiples 
of $\pi /2$, is the polyhedron necessarily orthogonal?  
The case of facial angles was answered previously 
in two papers at CCCG 2002 
\cite{DO-CCCG02,BCD+-CCCG02}: If a polyhedron with connected graph
has genus at most 2,
then facial angles that are multiples of $\pi/2$ imply that the
polyhedron is orthogonal, while for genus 6 or higher there exist
examples of non-orthogonal polyhedra with connected graph
where all facial angles are $\pi/2$.  

In this note we show that if both the facial and dihedral angles are multiples of $\pi /2$ 
then the polyhedron is orthogonal (presuming connectivity), 
and we give examples to show that the condition for dihedral 
angles alone does not suffice.

\section{Dihedral angles and facial angles}

We begin with the following question: Given a polyhedron
for which we know that every facial angle and every dihedral angle is
a multiple of $\pi/2$, is it an orthogonal polyhedron?  
This is not true if the graph of the polyhedron is disconnected
(see Figure~\ref{fig:counterex}), but it is true if the graph is
connected.  This can be seen as follows:  
Because the graph is connected, each face is simply connected, i.e., 
a polygon without holes.  Any simple connected polygon whose angles 
are all multiples 
of $\pi /2$ is necessarily orthogonal.  Rotate the polyhedron so that 
one face $f$ lies in a plane perpendicular to a coordinate axis with 
its edges parallel to coordinate axes.  Because the dihedral angles 
are multiples of $\pi /2$ therefore any face adjacent to $f$ lies in a 
plane perpendicular to a coordinate axis, and because the face is an 
orthogonal polygon, all  its edges are parallel to coordinate axes.   
Continuing to propagate to adjacent faces shows that the polyhedron is orthogonal.

\section{Dihedral angles only}

We now consider the following question:  If all dihedral angles are
multiples of $\pi/2$, is the polyhedron necessarily orthogonal, at
least if the graph is connected and the genus is small?  
The answer is ``no'', even for genus 0.  We can
even impose additional restrictions, such as all vertices having degree 3
or 4 and all faces being quadrangles.

\begin{figure}[ht]
\hspace*{\fill}
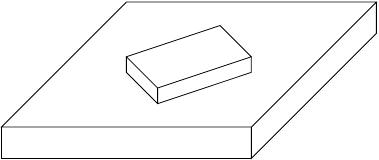
\hspace*{\fill}
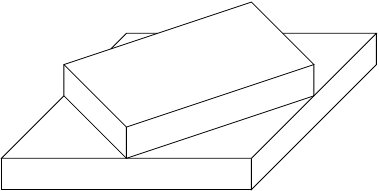
\hspace*{\fill}
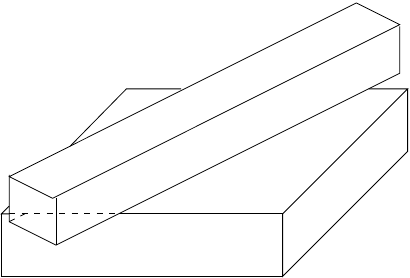
\hspace*{\fill}
\caption{Polyhedra where all dihedral angles are $\pi/2$ or $3\pi/2$.
(Left) The graph is disconnected.  Note that all facial angles are 
also $\pi/2$ or $3\pi/2$.  (Middle) Increasing the ``top'' makes the
graph connected.
Some vertices have degree 5.  (Right)  All vertices have degree 3 or 4,
all faces are quadrangles.}
\label{fig:counterex}
\end{figure}

Notice that the graph of the polyhedron in Figure~\ref{fig:counterex}(middle) could
not possibly be the graph of an orthogonal polyhedron (because it has a vertex
of degree 5), whereas the graph of the polyhedron in Figure~\ref{fig:counterex}(right)
can also be realized by
an orthogonal polyhedron (if we ``rotate'' the top part.) 
One may wonder how difficult it is to test whether a connected graph, together
with dihedral angles, forms an orthogonal polyhedron.  We can answer this in
the special case where we additionally know the lengths of the
edges, the graph is planar (i.e., the polyhedron has genus 0), and
no dihedral angle is $\pi$.  In this case, an orthogonal polyhedron, if one
exists at all, is unique and can be found in linear time \cite{BG11b}.
Hence we can run the algorithm to find it, and if it fails, conclude that
no orthogonal polyhedron can realize this graph, edge lengths and dihedral
angles.  In all other cases
(if edge lengths are unknown, dihedral angles may be $\pi$, or
the graph has higher genus) the complexity of testing realizability by
an orthogonal polyhedron remains open.

\bibliographystyle{alpha}
\bibliography{../bib/full,../bib/papers,../bib/gd}

\end{document}